\newcommand{\ignore}[1]{}
\documentstyle[11pt,epsf]{article}
\newcommand\be{\begin{equation}}
\newcommand\ee{\end{equation}}
\newcommand\bea{\begin{eqnarray}}
\newcommand\eea{\end{eqnarray}}
\newcommand\half{{\textstyle{1\over2}}}

\def\bmat{      \left |  \begin{array}{cc} }
\def\emat{ \end{array} \right |    }

\begin{document}
%\draft
%\title{}
%\author{}
%\address{}
%\date{\today}
%\maketitle
%\begin{abstract}
%\end{abstract}
%\pacs{}

\begin{titlepage}
\noindent BROWN-HET-1082  \hfill June 6, 1997 \\
\begin{center}

{\Large\bf  Diffractive Production at Collider
Energies}

{\Large\bf I: Soft Diffraction and Dino's Paradox}

\vspace{1.5cm}
{\bf Chung-I Tan$^{(1)}$}\\
\end{center}
\vspace{1.0cm}
\begin{flushleft}
{}~~$^{(1)}$Department of Physics, Brown University, Providence, RI 02912,
USA\\
\end{flushleft}
\vspace{1.5cm}

%Eq.~(\ref{eq:global}),
%~\cite{Peskin}
%Eq.~(\ref{eq:global}),

\abstract{One of the more interesting developments  from
recent collider experiments 
 is the finding that  hadronic  total cross sections as well as  elastic cross sections can be  described by
the exchange of a ``soft Pomeron" pole in the near-forward limit. The most important consequence of Pomeron
being a pole is the factorization property.
 However, due to 
Pomeron intercept being greater than 1,  the extrapolated 
single diffraction dissociation cross section based on a classical triple-Pomeron formula is too
large leading to a potential unitarity violation at Tevatron energies.  
 It is
our desire here to examine carefully the  notion of  Pomeron
flux renormalization, proposed by Goulianos, and to suggest an alternative resolution.  Our
treatment  lies in a proper implementation of final-state screening correction, with ``flavoring"
for Pomeron as the primary dynamical mechanism for setting the relevant energy scale. In our scheme,
which we shall occasionally refer to as ``unitarization of Pomeron flux", initial-state screening
remains unimportant, consistent with the pole dominance picture for elastic and total cross section
hypothesis at Tevatron energies.   Since factorization remains intact, umambiguous predictions for
double Pomeron exchange, doubly diffraction dissociation, etc., both at Tevatron and at LHC
energies, can be made.   }

\end{titlepage}

\section{Introduction}
One of the more interesting developments  from
recent collider experiments 
 is the finding that  hadronic  total cross sections as well as  elastic cross sections in the
near-forward limit can be  described by the exchange of a ``soft Pomeron"
pole,~\cite{Tan1} {\it i.e.},
 the absorptive part of the elastic amplitudes can be approximated by 
${Im}\> T_{a,b}(s,t)\simeq \beta_a(t) s^{\alpha_{\cal P}(t)}\beta_b(t).$
 The Pomeron trajectory has two important features.~\cite{PomeronFits} First, its
zero-energy intercept is greater than one,
$\alpha_{\cal P}(0)\equiv 1+
\epsilon$,
$\epsilon\simeq 0.08\sim 0.12$, leading to rising $\sigma^{tot}(s)$. Second, its Regge
slope is approximately $\alpha_{\cal P}'\simeq 0.25\sim 0.3$ $
GeV^{-2}$, leading to  
the observed  shrinkage effect for  elastic peaks. 
The most important consequence of Pomeron
being a pole is  factorization. For a singly diffractive dissociation process,
factorization leads to a  ``classical triple-Pomeron" formula,~\cite{Classical} 
\be
{d\sigma \over dtd\xi}\rightarrow {d\sigma^{classical} \over dtd\xi}\equiv  F_{{\cal P}/a}^{cl}
(\xi, t)
\sigma_{{\cal P}b}^{cl} (M^2,t),
\label{eq:ClassicalTP}
\ee
where $M^2$ is the missing mass variable and $\xi\equiv M^2/s$.   The first term on the right-hand
side of Eq. (\ref{eq:ClassicalTP}) is the so-called ``Pomeron flux", and the second term is the
``Pomeron-particle" total cross section.
Eq. (\ref{eq:ClassicalTP}) is in principle valid only when 
$\xi^{-1}$ and $M^2$ are both large,  with
$t$ small and held fixed.
However, with  $\epsilon\sim 0.1$, it has been observed~\cite{Dino1} that  the extrapolated
$p\bar p$ single diffraction dissociation cross section, $\sigma^{sd}$,  based on
the standard triple-Pomeron formula is too large at Tevatron energies by as much as a  factor of
$5\sim 10$ and it could   become larger than the
total cross section. 

Let us denote  the singly diffractive cross section  as a product
of a ``renormalization" factor and the classical formula,
\be
{d\sigma \over dtd\xi}=Z(\xi,t;s) {d\sigma^{classical} \over dtd\xi}.
\label{eq:TotalRenormalization}
\ee
It was  argued by K. Goulianos in Ref. \cite{Dino1} that
agreement with data could be achieved by having an energy-dependent suppression factor,
$Z(\xi,t; s)\rightarrow Z_{G}(s)\equiv  N(s)^{-1}\leq 1$. This ``renormalization" factor is chosen
so that the new ``Pomeron flux", 
{$F_N(s, \xi, t) \equiv N(s)^{-1}F_{{\cal P}/p}^{cl} (\xi, t)$}, is normalized  
 to unity for {$ s\geq \bar s$},{ $\sqrt{\bar s}\simeq
22\> GeV$}.~\cite{Dino1}\cite{Dino2}\cite{Dino3}   However,   the modified triple-Pomeron formula no
longer has a factorized form. An alternative suggestion has  been made recently by P.
Schlein.~\cite{Schlein1}\cite{Schlein2}   It was argued that phenomenologically, after incorporating
lower triple-Regge contributions,  the renomalization factor for the triple-Pomeron contribution
could be described by an
$\xi$-dependent suppression factor,
$Z\rightarrow Z_{S}(\xi)$.

 In view of the factorization property for total and elastic cross sections, the ``flux
renormalization"  procedure appears paradoxical and could undermine the theoretical foundation of a
soft Pomeron as a Regge pole from a non-perturbative QCD perspective.    We shall refer
 to this
as ``{\bf Dino's paradox}''.
 {Finding a resolution that is consistent with Pomeron
pole dominance for elastic and total cross sections at Tevatron energies will be the main focus of
this study.}  In particular, we want to   maintain the 
following factorization property,
\be
{ d\sigma \over
dtd\xi}\rightarrow \sum_k D_k(\xi,t)\Sigma_k(M^2),
\label{eq:FactorizedTR}
\ee
 when  $\xi^{-1}$ and $M^2$ become large.~\cite{Convention} 
It is
our intention  to provide  a solution without deviating from
generally accepted guiding principles for hadron dynamics. 

It is reasonable to expect that  the
resolution to the paradox should lie in a proper implementation of screening corrections to the
classical triple-Pomeron formula.  One of our key observations is the fact that, since $\sigma^{sd}$
is approximately $15\%$ of the total cross section at Tevatron, a pole dominance picture for
$\sigma^{tot}$ should also impose a  direct constraint on the nature of diffractive
production. 
The second
observation  is the necessity  in having an overall coherent scheme in which various different
hadronic energy scales must enter consistently. In our usage of classical triple-Regge formulas,
the basic energy scale is always in terms of ${s_0}\simeq 1$ $ GeV^2$. We demonstrate that there are
at least two other energy  scales, $s_r\equiv e^{y_r}$ and $s_f\equiv e^{y_f}$, 
$y_r\simeq 3\sim 5$ and $y_f\simeq  8\sim 10$,  which must be incorporated properly. The first is
associated with the physics of light quarks and the scale of chiral symmetry breaking.  The second
is the ``flavoring" scale and is associated with ``heavy flavor" production. In a non-perturbative
QCD setting, both play an important role in our understanding of a bare Pomeron with an intercept
greater than unity.~\cite{Flavoring4}\cite{Flavoring5}

 Our treatment  lies in a proper implementation of final-state screening correction, (or
{\bf final-state unitarization}), with ``{\bf flavoring}'' for Pomeron as the primary dynamical
mechanism for setting the relevant energy scale. In our treatment, initial-state screening remains
unimportant, consistent with the pole dominance picture for elastic and total cross section
hypothesis at Tevatron energies.  
The factor $D_{a,{\cal P}}(\xi,t)$ from the Pomeron
contribution will be referred to as a ``unitarized Pomeron flux factor,''   and  we
shall occassionally refer to our procedure as ``{\bf unitarization of Pomeron flux}''. 
We shall
demonstrate  that in our unitarization scheme the total renormalization  factor  has a 
factorized form, 
\be
Z(\xi,t;s)=Z_d(\xi,t)Z_m(M^2)=[S_f(\xi,t)R^2(\xi^{-1})]R(M^2),
\ee
where $S_f$ is due to final-state screening, Eq. (\ref{eq:FinalScreeningFactor}).  $R$ is a
flavoring factor, given by  Eq. (\ref{eq:FlavoringFactor}), and there is one flavoring factor
for each Pomeron propagator.

With the  pole dominance hypothesis, we demonstrate that  the
unitarized flux factor, $D_{a,{\cal P}}(\xi,t)\equiv Z_d(\xi,t)F^{cl}_{a,{\cal P}}(\xi,t)$, must
satisfy a normalization condition. 
\be
 \int_{-\infty}^0dt\int_{0}^{1}{d\xi} D_{a,{\cal P}}(\xi,t) \xi^{\epsilon} \equiv
\beta_a^{diff} < \beta_a(0).
\label{eq:SR1}
\ee
The required damping to overcome the $\xi^{-(1+2\epsilon)}$ divergent behavior of $F^{cl}_{a,{\cal
P}}(\xi,0)$ at small $\xi$  comes from both the screening factor
$S_f(\xi,t)$ and the factor $\xi^{\epsilon}$ from the ``Pomeron-particle'' total cross section. More
importantly, we point out that, even without   screening, flavoring alone would  lead to  a
suppression for the large 
$\xi^{-1}$ and large $M^2$ region  at Tevatron energies  of the
order $\{{R(\infty)/R(1)}\}^{3}\simeq e^{-3\epsilon y_f}\simeq  {1/ 8}$,
where $R(\infty)=1$. For instance,  flavoring  leads to a suppression  for the triple-Pomeron
coupling 
\be
g_{\cal PPP}(0)\sim e^{-3\epsilon y_f/2}g^{low}_{\cal PPP}(0),
\label{eq:TPSuppression1}
\ee
 where
$g^{low}_{\cal PPP}(0)\simeq 0.364\pm 0.025\>
mb^{1\over 2}$ is that obtained in a fit to low-energy data.~\cite{Dino5} In other
words, the apparent unitarity violation will be  mostly alleviated by flavoring alone! Lastly we
emphasize,   since factorization remains intact, our unitarization scheme leads to
unambiguous predictions  in terms of these suppression factors for double Pomeron exchange, doubly
diffraction dissociation, etc., both at Tevatron and at LHC energies.

In Section II, we explain why, given the Pomeron pole dominance hypothesis,  the initial-state
screening cannot be large at Tevatron energies. We introduce the notion of ``rapidity gap" cross
sections and explain  the sum rule, Eq. (\ref{eq:SR1}).  
 We study in
Sec. III the effect of final-state absorption. In particular, we demonstrate that, with Pomeron
intercept greater than one, absorption is ``total" within the ``expanding hadronic disk" and the
residual inelastic scattering can only take place    on the ``edge" of the disk.  It follows that
unitarity will not be violated by diffraction dissociation within a pole dominance picture for
$\sigma^{tot}$. A model  for calculating the screening factor, $S_f(\xi,t)$, is introduced. 
In Sec.
IV  we discuss the origin of the flavoring energy scale and indicate its relevant for diffractive
dissocition cross sections. In  a scale-dependent  formulation, we explain how flavoring contributes 
factors $R^2(\xi^{-1})$ and $R(M^2)$ to $Z(\xi,t;s)$ in order to account for both   the Pomeron
pole and  coupling renormalizations. The flavoring function  $R$ can be expressed in terms of a
scale-dependent  ``effective" Pomeron intercept, $\epsilon_{eff}(y)$.

Putting these together, we provide the {\bf final recipe} for our  resolution to  Dino's
paradox in Sec. V. Phenomenologically simple parametrizations for both the screening function,
$S_f$, and the effective intercept, $\epsilon_{eff}$,  are presented.  Readers  interested only in
our proposed modification to the classical triple-Pomeron formula can go there directly.   We
present a phenomenological analysis  which yields an estimate  for the ``high energy" triple-Pomeron
coupling: 
\be
g_{\cal PPP}(0)\simeq .14\sim .20 \>\>{mb}^{\half}. 
\ee
This value  is consistent with our
flavoring expectation, Eq. (\ref{eq:TPSuppression1}).  Surprisingly, the amount of screening
required at Tevatron energies seems to be very small. We  discuss our predictions for double Pomeron
exchange, double diffraction dissociation, and other multi-gap processes in Sec. VI. Comparison of
our approach to that of Refs.
\cite{Dino1} and
\cite{Schlein1} as well as other  comments are given in Sec. VII.

\section{ Pomeron Dominance Hypothesis at Tevatron Energies}

We shall  first explore consequences of the observation that both total cross sections and
elastic cross sections can be well described  by a Pomeron pole exchange at Tevatron energies.
Absorption correction, if required, seems to remain small.  Since the singly diffractive cross
section,
$\sigma^{sd}$, is a sizable part of the total, it must also grow as $s^\epsilon$. This
qualitative understanding can be quantified in terms of a sum rule for ``rapidity  gap" cross
sections. This in turn imposes a convergence condition on our unitarized Pomeron flux, $D_{a,{\cal
P}}(\xi,t)$.

To simplify the discussion, we shall first ignore transverse momentum distribution by treating the
longitudinal phase space only. For instance, for  singly diffraction dissociation, the
longitudinal phase space can be specified by two rapidities,
$y\equiv \log (\xi^{-1})$ and $y_m\equiv \log M^2.$
 The first variable specifies the rapidity
gap associated with the detected leading proton (or antiproton), and the second variable specifies
the rapidity ``span" of the missing mass distribution. At fixed $s$, they are constrained by
$y+y_m\simeq Y\equiv \log s$, (see Figure~\ref{fig:phase_space}), and we can speak of  differential diffractive cross
section
$d\sigma^{sd}/dy$. We shall in what follows use $\{\xi^{-1},\> M^2\}$ and $\{y\equiv
\log{\xi^{-1}}, \> y_m\equiv \log M^2\}$ interchangably.  Dependence on transverse degrees of freedom
can be re-introduced without much effort after completing the main discussion.

Consider  the process $a+b\rightarrow c+ X^{<}$, where
the number of particles in $X$ is  unspecified. However, unlike the usual single-particle
inclusive process, the superscript for
$X^{<}$ indicates that all particles in $X$ must have rapidity less than that of the particle $c$,
{\it i.e.}, the detected particle $c$ is the one in the final state with the largest  rapidity value. 
Kinematically, a single-gap cross section is  identical to the singly diffraction dissociation
cross section discussed earlier. Under the assumption
where all transverse motions are unimportant, one has $y_c\simeq y_a\equiv  Y_{max}$ and the 
differential gap cross section, 
${d\sigma^{gap}/dy}$, can also be considered as a function of $y$ and $y_m$, with $y+y_m\simeq
Y$.  

Because the detected particle $c$ has been singled out to be different from all other
particles, this is  no longer an inclusive cross section, and it does not satisfy
the usual inclusive sum rules.
Upon integrating over the rapidity gap
$y$ and summing over particle type $c$, no multiplicity enhancement factor is introduced and one
obtains simply the total cross section, {\it i.e.}, a gap cross section satisfies the following exact sum
rule,
\be
\sum_c \int dy {d\sigma_{ab\rightarrow c}^{gap}\over dy}\equiv \sum_c\sigma_{ab\rightarrow
c}^{gap} =
\sum_{n=2}^{\infty} \sigma_n= 
\sigma^{tot}_{ab}.
\label{eq:SumRule}
\ee
Interestingly, this allows  an
identification of the toal cross section as a sum over specific gap cross sections,
$\sigma_{ab\rightarrow c}^{gap}$, each is ``derived" from a specific ``leading particle" 
gap distribution.
 Since no restriction has been imposed on the nature  of
the ``gap distribution",
{\it e.g.}, particle
$c$ can have different quantum numbers from
$a$, the notion of a gap cross section is more general than a diffraction cross
section.\cite{Diffraction} It follows that the singly diffractive
dissociation cross section,
$\sigma^{sd}_{ab}$, is a part of $\sigma_{ab\rightarrow
a}^{gap}$.

Consider next our  factorized ansatz, Eq. (\ref {eq:FactorizedTR}). For $y$ and $y_m$ large, it
leads to a gap distribution, ${d\sigma_{ab\rightarrow c}^{gap}/dy}= e^{-y}D_{a\rightarrow
c}(y)\Sigma_b(y_m).$ 
If  $\Sigma_b(y_m)=\beta_b e^{\epsilon y_m}$, it follows that
contribution from each gap distribution is Regge behaved, $\sigma^{gap}_{ab\rightarrow c} \simeq
\beta_a^ce^{\epsilon Y}\beta_b$, where the total Pomeron residue is a sum of ``partial
residues", 
\be
\beta_a=\sum_c \beta_a^c=\sum_c\int_0^{\infty} dy D_a^c(y)e^{-\epsilon y},
\label {eq:PomeronResidue}
\ee
For above integral to converge, each flux factor must grow slower than $e^{\epsilon y}$. That
is, 
$D_a^c(y)e^{-\epsilon y}\rightarrow 0$ as $y\rightarrow \infty.$
In a traditional Regge approach, the large rapidity gap behavior for each flux factor is controlled
by an approriate  Regge propagator, $e^{(\alpha_i+\alpha_j-1)y}$. Clearly a standard triple-Pomeron
 behavior with $\alpha_{\cal P}>1$ is inconsistent with the pole dominance hypothesis.
Unitarity correction must supply enough damping to provide convergence.

 There is yet another 
way of expressing the consequence of  the pole dominance hypothesis. Dividing each gap differential
cross section by the total cross section,  factorization of Pomeron leads to a ``limiting 
distribution": $\rho_{ab}(y,Y)\rightarrow \rho_a(y)\equiv \sum_c \rho_a^c(y).$
That is, the limit is independent of the total rapidity, $Y$, and the gap density is
normalizable, $\int_0^{\infty}dy \rho_a(y)=\sum_c\int_0^{\infty}dy
\rho_a^c(y)=\sum_c{(\beta_a^c/ \beta_a)}=1$. This normalization condition for the gap
distribution is precisely Eq. (\ref{eq:PomeronResidue}).

Let us now restore the transverse distribution and concentrate on the  diffractive dissociation
contribution, which can be identified with the high $M^2$ and high $\xi^{-1}$ limit of
$d\sigma^{gap}_{ab\rightarrow a}(t,\xi;M^2)$. In terms of
$\xi$,
$M^2$, and $t$, the differential cross section at large $M^2$ under our factorizable ansatz  takes
on the following form, $
{d\sigma_{ab}^{sd}/ dt d\xi} \simeq D_{a,{\cal P}}(\xi,t) \Sigma^{cl}_{{\cal
P},b}(M^2),$ where $\Sigma^{cl}_{{\cal
P},b}(M^2)=(M^2)^{\epsilon}\beta_b(0).$
It follows from Eq. (\ref{eq:PomeronResidue}) that 
$D_{a,{\cal P}}(\xi,t)$ must satisfy the following bound:
\be
 \int_{-\infty}^0dt\int_{\xi_{min}(s)}^{\xi_{max}}{d\xi} D_{a,{\cal P}}(\xi,t) \xi^{\epsilon}\leq 
\int_{-\infty}^0dt\int_{0}^{1}{d\xi} D_{a,{\cal P}}(\xi,t) \xi^{\epsilon} \equiv
\beta_a^{diff} < \beta_a(0).
\label{eq:UnitarizedTPSumRule}
\ee

The hypothesis  of a Pomeron pole dominance for the total and elastic cross sections is of course
only approximate. However, to the extend that absorptive corrections remain small at Tevatron
energies, one finds that  a modified Pomeron flux factor  must
differ from the ``classical" Pomeron flux at small $\xi$ in such a way so that the
upper bound in Eq. (\ref{eq:UnitarizedTPSumRule}) is satisfied.
 We shall refer to $D_{a,{\cal P}}(\xi,t)$ as the ``unitarized Pomeron  flux".  How this can be 
accomplished via final state screening will be
discussed next.
 Note both the
{\bf similarity} and the {\bf  difference} between Eq. (\ref{eq:UnitarizedTPSumRule}) and the ``flux
normalization" condition mentioned in the Introduction. Here, this convergent integral yields a
finite number,
$\beta^{diff}_a$, and  the ratio
$\beta^{diff}_a/\beta_a(0)$ can be interpreted as the probability of having a diffractive gap at
high energies.

\section{Final-State Screening}

The best known example for implementing the idea of ``screening" in high energy hadronic collisions
has been the ``expanding disk" picture for rising total cross sections. Diffraction scattering
as the shadow of inelastic production has been a well established mechanism for the occurence
of a forward peak. Analyses of data upto collider energies have revealed that the essential feature
of nondiffractive particle production  can be understood in terms of a multipertipheral
cluster-production mechanism. In such a picture, the forward amplitude is predominantly absorptive
and is dominated by the exchange of a ``bare Pomeron".  If the Pomeron intercept is greater than
one, it forces further unitarity corrections as one moves to higher energies. For instance,
saturation of the Froiossart bound can be next understood through an eikonel mechanism, with the
absorptive eikonal
$\chi(s,b)$ given by the bare Pomeron amplitude in the impact-parameter space.

 The main problem we are facing
here is not so much on how to obtain a ``most accuraate" flux factor $D_{a,{\cal P}}(\xi,t)$ at
very small $\xi$.  We are concerned with 
a more difficult conceptual problem of how to reconcile having a potentially large screening
effect for diffraction dissociation processes and yet being able to maintain approximate pole
dominance for elastic and total cross sections up to Tevatron energies.  We shall show using an
expanding disk picture that absorption works in such a way that inelastic scattering
can only take place on the ``edge" of disk.~\cite{EdgeEffect} Therefore, once applied using
final-state screening, the effect of initial-state absorption will be small, hence allowing us to
maintain Pomeron pole factorization for elastic and total cross sections. 

\subsection{Expanding Disk Picture}

Let us briefly review this
picture which also serves to establish notations. At high energies, a near-forward amplitude can be expressed in an impact-parameter representaion via a two-dimensional Fourier
transform,
\be
T(s,t)\equiv
2is\int d^2{\vec b} e^{i\vec q\cdot\vec b} \tilde f(s,b),
\hskip30pt  
\tilde f(s,b)= (4i\pi s)^{-1} \int d^2\vec q e^{-i\vec b\cdot\vec q} T(s,t),
\ee
where $t\simeq -\vec q^2$.  Assume that the near-forward elastic amplitude at
moderate energies can be described by a Born term, {\it e.g.}, that given by a single Pomeron
exchange where we shall approximate it to be purely absorptive.
 Let us  denote the contribution from  the Pomeron exchange to $\tilde f(s,b)$ as $\chi(s,b)$.
With $\alpha_{\cal P}(t)=1+\epsilon+\alpha'_{\cal P}  t$, and approximating $\beta(t)$ by an
exponential, we  find $\chi (s,b)\simeq X(s) e^{-(b^2/ 4 B(s))}$, 
\be
 X(s)\equiv
\chi(s,0)\simeq {\sigma_{0}(s)\over 4B(s)}
\ee
where 
$B(s)=b_0+\alpha'_{\cal P} \log s$
and $\sigma_{0}(s)= \sigma_0 s^{\epsilon}$.
%\{$ Will eventually use $\epsilon\simeq 0.08$, $b_0\simeq 4.6$ and $\alpha'_{\cal P}=.26$.
%For $\sigma_0=(14.3 /.389) GeV^{-2})$. 
%This corresponds to $\sigma(s) =(14.3 /.389) s^{\epsilon}$ in $GeV^{-2}$.   $\}$
With
$\epsilon>0$, the Born approximation would eventually violate $s$-channel unitarity  at small $b$ as 
$s$ increases. A systematic procedure which in principle should restore unitarity is the  Reggeon
calculus. However, our current understanding of dispersion-unitarity-relation is
too qualitative to provide a definitive  calculational scheme.

 The key  ingredient of ``screening" correction is  the recognition that the next order
correction to the Born term
 must have a negative sign. (The sign of double-Pomeron cut contribution.) In an impact-representation,
Reggeon calculus assures us that the correction can be represented as 
\be
\tilde f_1(s,b) \simeq  -{1\over 2!}\> \mu(s)\>  \chi(s,b)^2,
\label {eq:TwoReggeonCut}
\ee
where $\mu$ is  positive. To go beyond this, one needs a model. 
A physically well-motivated model which should be
meaningful  at moderate energies and allows easy analytic treatment is the eikonal model. 
Writing  $ \tilde f(s,b)=\tilde f_0(s,b)+\tilde f_1(s,b) +\tilde f_2(s,b)+\cdots,$ the expansion alternates in
sign, and with simple weights such that
$\tilde f(s,b)= [1-e^{-\mu \chi(s,b)}]/\mu $, 
and 
\be
T(s,t)=({2 i s \over \mu}) \int d^2{\vec b} \> e^{i\vec q\cdot\vec b}\{1-e^{-\mu \chi(s,b)}\}.
\ee
{Conventional eikonal model has $\mu=1$. We keep
$\mu\leq 1$ here so as to allow the possibility that screening is ``imperfect".} 

Observe that the  eikonal derived from the Pomeron exchange, $\chi(s,b)$, is
a monotonically decreasing function of
$b^2$, taking on its maximum value $X(s)$ at $b^2=0$, 
which  increases with $s$ due to $\epsilon>0$. The eikonal drops to zero  at
large $b^2$ and  is of the order $1$ at a radius, $b_c(s)\simeq \sqrt
{B(s)\log \mu X(s)}\sim \log s.$  Within this radius, $\tilde f(s,b)=O(1)$ and it
vanishes beyond. This is the ``expansion disk" picture of high energy scattering, leading to an
asymptotic  total  cross section
$O(b_c(s)^2)$.

\subsection{Inelastic Screening}

In order to discuss inelastic final-state screening, we follow the ``shadow" scattering picture in
which the ``minimum biased" events are predominantly ``short-range ordered" in rapidity and the
production amplitudes can be described by  a multiperipheral cluster model. Substituting
these into the right-hand side of an elastic unitary equation, $Im T(s,0)= \sum_n |T_{2,n}|^2$,
one finds that the resulting elastic amplitude is 
dominated by the exchange of a Regge pole, which we shall provisionally refer to as the
``bare Pomeron".
Next consider singly
diffractive events. We assume that the ``missing mass" component corresponds to  no gap events, thus the distribution is again
represented by a  ``bare Pomeron". However, for the gap distribution, one would insert the ``bare
Pomeron" just generated into a production amplitude, thus leading to the classical triple-Pomeron
formula.

Extension of this scheme leads to a ``perturbative" treatment for the total cross
section in the number  of bare Pomeron exchanges along a multiperipheral chain.  Such a scheme was
proposed long time ago,~{\cite{FrazerTan}} with the understanding that the picture could make sense
at moderate energies, provided that the the intercept of the Pomeron is one,
$\alpha_{\cal}(0)\simeq 1$, or less. However, with the acceptance  of a Pomeron having an intercept
greater than unity, this expansion must be embellished. Although it is still meaningful to have a
gap expansion, one must improve the descriptions for parts of a production amplitude  involving large
rapidity gaps by taking into account   absorptions for the gap distribution. We propose that this
``partial unitarization" be done for each gap separately, thus maintaining the factorization property
along each short-range ordered sector. This involves final-state screening, and, for singly
diffraction dissociation, it corresponds to the inclusion of ``enhanced Pomeron diagrams" in
the triple-Regge region.

 To simplify the notation, we shall use  the energy variable, $\xi^{-1}$,
and the rapidity gap variable, 
$y=\log(\xi^{-1})$, interchangeably. Let's express the total unitarized contribution to a gap cross
section in terms of a ``unitarized flux"  factor, $D(y,t) \equiv (e^y/ 16\pi)  |g_d(y ,
t)|^2\equiv (1/16\pi\xi)  |f_d(\xi , t)|^2{g_{\cal PPP}(t)},$
so that it reduces to the classical triple-Pomeron formula as its Born term. That is, the corresponding Born amplitude  for
$f_d(\xi, t)$ is the ``square-root" of the triple-Pomeron contribution to the classical formula, 
$f^{cl}_{\cal P}(\xi  ,t)= \beta_{\cal P}(t) (\xi^{-1})^{(\alpha_{\cal P}(t)-1)}.
$
Screening becomes important if  large gap becomes favored, {\it i.e.,} when $\epsilon>0$.  

Let us work in an
impact representation, $ g(y_d ,t)\equiv  {2i } \int d^2\vec q e^{i\vec b\cdot\vec q} \tilde g_d(y_d
,b).
$
Consider an  expansion
$
\tilde g(y ,b)=\chi_d(y,b)
+\tilde g_{1}(y,b) +\tilde g_{2}(y,b)+\cdots
$
where, under the usual exponential approximation for the $t$-dependence, the Fourier transform for
the Born term is $\chi_d(y ,b) 
={\sigma^{d}(y)\over 4B_d(y)} e^{-{b^2\over 4 B_d(y)}}$, 
with  
$B_d(y)=b_d+\alpha'_{\cal P} y$
and $\sigma^{d}(y)= \sigma_d e^{\epsilon\> y}$.
The key physics of absorption is 
\be
\tilde g_{1}(y ,b)\simeq - \mu_d \chi(y,b)\chi_d(y,b).
\ee
The proportionality constant  $\mu_d$ can be different from the constant
$\mu$ introduced earlier for the elastic screening, either for kinematic or dynamic reasons, or
both. For a generalized eikonal approximation, one has 
$\tilde g(y,b)= \chi_d  \{1- \mu_d \chi+ {1\over 2!}(\mu_d \chi)^2 -\cdots\}=\chi_d(y,b)
e^{\>-\mu_d \>\chi(y,b)}.$
If we define the ``final-state screening" factor as the ratio between the unitarized flux factor
and the classical triple-Pomeron formula, 
$
D_{a,{\cal
P}}(y,t)=S^{(0)}_f(y,t;X) F^{cl}_{a,{\cal
P}}(y,t), 
$
we then have
\be
S^{(0)}_f(y,t;X)= |f_d(y, t)/f_d^{cl}(y , t)|^2.
\label{eq:ScreeningFactor}
\ee
  We shall use this
expression  as a model for probing the physics of inelastic screening in an
expanding disk picture.

Let us examine this eikonal screening factor in the forward limit, 
$t=0$, where 
\be
S^{(0)}_f(y,0;X)^{1\over 2}={\int d{b^2} \chi_d(y,b)e^{-\mu_d \chi(y,b)}\over \int d{b^2}
\chi_d(y_d,b)}.
\ee
  Unlike the elastic
situation, the integrand of the numerator is strongly suppressed both in the region of large $b^2$
and in the region ``inside" the expanding disk.  {\bf The only significant  contribution comes from a
``ring" region near the edge of the expanding disk.}~\cite{EdgeEffect}\cite{InitialState} Since the
value of the integrand is of $O(1)$ there, one finds that the numerator varies with
energy only weakly. On the other hand, the denominator is simply
$\sigma^d(y)$, which increases as
$ e^{\epsilon y}$. Therefore, this  leads to an exponential cutoff in $y$,
$S^{(0)}_f(y,0;X)\>\sim\>  e^{-2\epsilon y}.$ This damping factor precisely cancels the
$\xi^{-2\epsilon}$ behavior from the classical triple-Pomeron formula at small $\xi$, leading to a
unitarized Pomeron flux factor.

 To be more precise, let us work in a simpler representation for $S^{(0)}_f(y,0;X)$ by
changing   the variable
$b^2\rightarrow z\equiv e^{-b^2/4B(y)}$. 
 With our gaussian
approximation, one finds that 
\be
S^{(0)}_f(y,0;X)=\bigl\{r\int_0^1dz z^{r-1}e^{-x z}\bigr\}^2|_{{x=\mu_dX(y)},r=r(y)}, 
\label{eq:ScreeningTzero}
\ee
where $r(y)\equiv B(y)/B_d(y)$. This expression can be expressed as
$\bigl\{rx^{-r}\Gamma(x,r)\bigr\}^2|_{{x=\mu_dX(y)},r=r(y)}$, where $\Gamma(x, r)=\int _0^{x}d z
z^{r-1} e^{-z}$ is  the  incomplete Gamma function.  In this representation, one easily verifies
that the screening factor has the desired properties: As
$\mu_d
\rightarrow 0$, screening is minimal and one has $S^{(0)}_f(y,0;X)\rightarrow 1$. On the
other hand, for
$y$ large, $X(y)$ increases so that 
$S^{(0)}_f(y,0;X)\rightarrow [\mu_d X(y)] ^{-2}$, as anticipated.

%S^{(0)}_f(y,0;\epsilon)=r(y)^2[\mu_d X(y;\epsilon )] ^{-2r(y)}
%{\Gamma(r(y),\mu_d X(y;\epsilon))}^2,

Similarly, we find that the logrithmic width for the unitarized flux $D(y,t)$ at $t=0$ has
increased from $2B_d$ to
$2B^{eikonal}_d$ where 
$B^{eikonal}_d=B_d\{ -r{d\over d r} \log {\int_0^1 d z z^{r-1}e^{-x
z}}\}_{{x=\mu_dX(y)},r=r(y)}
$.
As $\mu_d \rightarrow 0$, $B^{eikonal}_d\rightarrow B_d$. For $y$ very large,
$B^{eikonal}_d\rightarrow B\log\mu_d X(y)\sim b_c(y)^2\propto y^2.$  This corresponds to a
faster shrinkage than that of ordinary Regge behavior.
Averaging  over $t$, one finds at large diffractive rapidity $y$, the final-state screening
provides an avearge damping 
\be
\langle S_f \rangle \rightarrow e^{-2 \epsilon\> y}=\xi^{2\epsilon}.
\ee
This leads to a unitarized Pomeron flux, $D_{a,{\cal P}}(\xi,t)$,  which automatically satisfies the
upper bound, Eq. (\ref{eq:UnitarizedTPSumRule}), derived  from the Pomeron pole dominance
hypothesis.

%Will eventually use $\epsilon\simeq 0.08$,
%$b_d\simeq 4.6/2$ and $\alpha'_{\cal P}=.26$.

\newpage

\section{Dynamics for Soft Pomeron and Flavoring}

Although final-state screening would automatically avoid unitarity
violation, the primary source of high energy suppression actually comes from a proper treatment of
{\bf scale-dependence} for Pomeron couplings.  Consider for
the moment the following scenario where one has two different fits to hadronic total cross sections:
\begin{itemize}
\item
(a) ``High energy fit": $\sigma_{ab}(y)\simeq \beta_a\>\beta_b \>e^{\epsilon\> y}$ \hskip50pt for
\hskip50pt
$Y>>y_f$,
\item
(b) ``Low energy fit": $\sigma_{ab}(y)\simeq \beta^{low}_a\>\beta_b^{low} $ \hskip50pt for
\hskip40pt $0<Y<<y_f$.
\end{itemize}
That is, we envisage a situation where the ``effective Pomeron intercept", $\epsilon_{eff}$, 
changes as one moves up in energies.   Assuming a
smooth interpolation between these two fits, one can obtain the following order of magnitude estimate
\be
\beta_p\simeq e^{-{\epsilon\> y_f\over 2}} \beta_p^{low}.
\label{eq:HiLoResidue}
\ee
 Modern parametrization for Pomeron residues typically
leads to values of the order $(\beta_p)^2\simeq  14\sim 17 $ mb. However, before the advent of the
notion of a Pomeron with an intercept greater than 1, a typical parametrization would have a value
$(\beta^{low}_p)^2\simeq 35\sim 40$ mb, accounting for a near constant Pomeron contribution at low
energies. This leads to an estimate of $y_f\sim 8$, corresponding to $ \sqrt s \sim 50$ GeV.
This is precisely the energy scale where a rising total cross section first becomes noticeable.

The scenario just described has been referred to as ``flavoring", the notion that the
underlying effective degrees of freedom for Pomeron will increase as one moves to higher
energies,~\cite{Flavoring1}\cite{Flavoring2}\cite{Flavoring3}   and it has provided a dynamical basis
for understanding the value of Pomeron intercept in a non-perturbative QCD
setting.~\cite{Flavoring4}\cite{Flavoring5} In this scheme, both the Pomeron intercept and the
Pomeron residues are {\bf scale-dependent}. We shall briefly review this mechanism and introduce a
scale-dependent formalism where  the entire flavoring effect can be absorbed into a flavoring factor,
$R(y)$, associated with each Pomeron propagator. 

\subsection{Bare Pomeron in Non-Perturbative QCD}

In a non-perturbative QCD setting, the Pomeron intercept is
directly related to the strength of the short-range order component of inelastic production and this
can best be understood in a large-$N$ expansion.~\cite{LeeVeneziano}\cite{DPM}  In such a scheme,
particle production mostly involves emitting ``massless pions", and  the basic energy scale of
interactions is that of ordinary vector mesons, of the order of 
$1$ GeV.  In a one-dimensional multiperipheral realization for the ``planar
component" of the large-$N$ QCD expansion, the high energy behavior of a $n$-particle total cross
section is primarily controlled by its longitudinal phase space,
$\sigma_n\simeq (g^4N^2/{(n-2)!})(g^2N\log s)^{n-2} s^{J_{eff}-1}.$
Since there are only Reggeons at the planar diagram level, one has  $J_{eff}=2\alpha_R-1$ and, after
 summing over $n$,  one  arrives at Regge behavior for the planar component of $\sigma^{tot}$ where 
\be
\alpha_R=(2\alpha_R-1)+g^2N.
\label{eq:Planar}
\ee
At next level of cylinder topology,  the contribution to partial cross section increases due to its topological twists,
$\sigma_n\simeq {( g^4/ {(n-2)!})} 2^{n-2}(g^2N\log s)^{n-2} s^{J_{eff}-1},$
and, upon summing over $n$,  one arrives at a total cross section governed by a Pomeron exhange, 
\be
\sigma_0^{tot}(Y)=g^4e^{\alpha_{\cal P} Y}
\label {eq:BarePomeron}
\ee
where the  Pomeron interecept is 
\be
\alpha_{\cal P}=(2\alpha_R-1)+2g^2N.
\label{eq:Cylinder}
\ee
Combining Eq. ({\ref{eq:Planar}) and Eq. (\ref{eq:Cylinder}), we arrive at the following
amazing ``bootstrap" result,
\be
\alpha_{\cal P}\simeq 1.
\label{eq:Topology}
\ee

In a non-perturbative QCD setting, having a Pomeron intercept near 1 therefore depends crucially on
the topological structure of large-$N$ non-Abelian gauge
theories.~\cite{Flavoring4}\cite{Flavoring5}\cite{LeeVeneziano}\cite{DPM}
 In this picture, one has  $\alpha_R\simeq .5\sim .7$ and  $g^2N\simeq .3\sim .5  $.  With
$\alpha'\simeq 1$ $GeV^{-2}$, one can  also   directly relate 
$\alpha_R$ to the  average mass of typical vector mesons.
Since vector meson masses are controlled by constituent mass for light
quarks, and since constituent quark mass is a consequence of chiral symmetry breaking, the 
Pomeron and the Reggeon intercepts are directly related to fundamental issues in non-perturbative 
QCD.

Finally we note that, in a  Regge
expansion, the relative importance  of secondary trajectories  to the Pomeron is controlled by the
ratio
$e^{\alpha_{R}\> y}/e^{\alpha_{\cal P}\> y}= e^{-(\alpha_{\cal P}-\alpha_{\cal R})\> y}$. It follows 
that the relevant scale in rapidity we are seeking is precisely 
$(\alpha_{\cal P}-\alpha_R)^{-1}<y_r \simeq 3\sim 5$.
The importance  of this scale $y_r$ is of course well known:  When using a
Regge expansion for total and two-boby cross sections, secondary trajactory contributions become
important and must be included whenever rapidity separations are  below $3\sim 5$
units. This is of course also true for the triple-Regge region. That is, in
terms of $y_d$ and $ y_m$,  triple-Regge terms other than the leading triple-Pomeron term must be
added whenever one moves into regions when $y_d\leq y_r$ or $y_m\leq y_r$ or
both.~\cite{Schlein1}\cite{Schlein2}

\subsection{Flavoring of Bare Pomeron}

We have argued previously that ``heavy flavor" production provides an additional
energy scale, $s_f=e^{y_f}$,  for soft Pomeron
dynamics,  and the  effect of 
heavy flavors
can be  responsible for the perturbative increase of the Pomeron intercept to be greater than unity, 
$\alpha_{\cal P}(0)\sim
1+\epsilon,\>\>\epsilon>0$. One must bear this additional energy scale in mind in working with a soft
Pomeron.~\cite{Flavoring4}\cite{Flavoring5}\cite{Flavoring1}\cite{Flavoring2}\cite{Flavoring3}
That is, to fully justify using a Pomeron with an intercept $\alpha_{\cal P}(0)>1$, one must restrict
oneself to energies
$s>s_f$ where heavy flavor production is no longer suppressed. Conversely, to extrapolate Pomeron
exchange to low energies below $s_f$, a lowered ``effective trajectory" must be
used. This feature of course is
unimportant for total and elastic cross sections at Tevatron energies. However, it is important for
diffractive production since both $\xi^{-1}$ and $M^2$ will sweep right through this energy  scale at
Tevatron energies. (See Figure~\ref{fig:phase_space}.)

Flavoring becomes important whenever  there is a further inclusion of effective degrees of freedom
than that associated with light quarks. This can again be illustrated by a simple one-dimensional
multiperipheral model. In addition to what is already contained in the Lee-Veneziano
model, suppose that new particles can also be produced in a multiperipheral
chain. Concentrating on the cylinder level, the partial  cross sections will be  labelled by
two indices, 
\be
\sigma_{p,q}\simeq  (g^4/ p!q!) 2^{p+q}(g^2N\log s)^{p} (g^2_{f}N\log
s)^{q}s^{J_{eff}-1},
\label{eq:FlavoringTotalCrossSection}
\ee
where $q$ denotes the number of clusters of new particles produced. Upon summing over $p$ and
$q$, we obtain a ``renormalized" Pomeron trajectory
\be 
\alpha_{\cal P}=\alpha^{old}_{\cal P}+ \epsilon,
\label{eq:NewPomeron}
\ee
where $\alpha^{old}_{\cal P}\simeq 1$ and $\epsilon\simeq  2g^2_{f}N$. That is, in a non-perturbative
QCD setting, the intercept of Pomeron is a dynamical quantity,  reflecting the effective degrees
of freedom involved in near-forward particle
production.\cite{Flavoring4}\cite{Flavoring5}\cite{Flavoring6}

If  the new degree of freedom involves
particle production with high mass, the longitudinal phase space factor, instead of $({\log
s})^q$, must be modified. Consider the situation of producing one $Q\bar Q$ bound state together
with pions, {\it i.e.}, $p$ arbitrary and  $q=1$ in Eq. (\ref{eq:FlavoringTotalCrossSection}). 
Instead of
$(\log s)^{p+1}$,  each factor  should be replaced by
$(\log(s/m^2_{eff}))^{p+1}$, where $m_{eff}$ is an effective  mass for the $Q\bar Q$ cluster.
In terms of rapidity, the longitudinal phase space factor becomes
$(Y-\delta)^{p+1}$,  where
$\delta$ can be thought of as a one-dimensional ``excluded volume" effect. 
For heavy particle production, there will be an energy range over which $q=1$ remains a valid
approximation. Upon summing over $p$, one finds that the additional contribution to the total cross
section due to  the production  of one heavy-particle cluster is~\cite{Flavoring1}\cite{Flavoring2} 
\be
\sigma^{tot}_{q=1}\sim \sigma_0^{total}(Y-\delta)(2g^2_fN)\log
(Y-\delta)\theta(Y-\delta),
\label{eq:OneParticleFlavoring}
\ee
where $\alpha_{\cal P}^{old}\simeq 1$. 

Note the effective longitudinal phase space ``threshold
factor",
$\theta(Y-\delta)$, and, initially, this term  represents a small perturbation  to the total
cross section obtained previously, (corresponding to
$q=0$ in Eq. (\ref{eq:FlavoringTotalCrossSection})), $\sigma_0^{total}$. 
Over a rapidity range, $[\delta, \delta+\delta_f]$, where $\delta_f$ is the average rapidity
required for producing another heavy-mass cluster, this is  the only term needed for incorporating
this new degree of freedom. As one moves to higher energies,
``longitudinal phase space suppression" becomes less important and more and more heavy particle
clusters will be produced. Upon summing over
$q$, we would obtain a new total cross section, described by  a  renormalized Pomeron,  with a new
intercept given by Eq. (\ref{eq:NewPomeron}).

We  assume that, at Tevatron, the energy is high enough so that this kind of ``threshold"
effects is no longer important.  How low an energy do we have to go before one encounter these
effects? Let us try to answer this question by starting out from low energies. As we have stated
earlier, for
$Y> 3\sim 5$, secondary trajectories become unimportant and using  a Pomeron with
$\alpha_{\cal}\simeq 1$ becomes a useful approximation. However, as new flavor production becomes
effective, the Pomeron trajectory will have to be renormalized. We can estimate for the relevant
rapidity range when this becomes important as follows:  $y_f >  2
\delta_{0}+
<q>_{min} \delta_f$. The first factor
$\delta_{0}$ is associated with leading particle effect, i.e., for proton, this is primariy due
to pion exchange.
$\delta_f$ is the minimum gap associated with one heavy-mass cluster  production, {\it  e.g.},
nucleon-antinuceon pair production. We estimate
$\delta_{0}\simeq  2$ and $
\delta_f\simeq 2\sim 3 $,  so that, with $<q>_{min}\simeq 2$,  we  expect the
relevant flavoring rapidity scale to be 
$y_f\simeq 8\sim 10$.

\subsection {Effective intercept  and Scale-Dependent  Treatment}

In order to be able to extend  a Pomeron repesentation below the rapidity scale $y\sim y_f$, we
propose   the following {\bf scale-dependent}  scheme where we introduce a flavoring factor for each
Pomeron propagator.  Since each
Pomeron exchange is always associated with  energy variable
$s$, (therefore a rapidity variable
$y\equiv
\log s$), we shall  parametrize the Pomeron trajectory function as 
\be
\alpha_{eff}(t; y)\simeq 1+\epsilon_{eff}(y) +\alpha' t,
\label{eq: EffectivePomeronTrajectory}
\ee
where $\epsilon_{eff}(y)$ has the properties
\begin{itemize}
\item 
{(i)} $\epsilon_{eff}\simeq \epsilon_o\equiv \alpha^{old}_{\cal P}-1\simeq 0 $ \hskip55pt for
\hskip40pt 
$y<< y_f$,
\item 
{(ii)}  $\epsilon_{eff}\simeq \epsilon \simeq 0.1  $ \hskip100pt for \hskip40pt  $y>>y_f$.
\end{itemize}
For instance, exchanging such an effective Pomeron leads to a contribution to the elastic cross
section 
\be
T_{ab}(s,t)\propto s^{1+\epsilon_{eff}(y) +\alpha' t}.
\ee
This representation  can now be extended down to the region $y\sim y_r$. 
We shall adopt a particularly convenient parametrization for $\epsilon_{eff}(y)$ in the next
Section when we discuss phenomenological concerns.

To complete the story, we need also to account for the scale dependence of Pomeron residues. What we
need is an ``interpolating" formula between the high energy and low energy sets. Once a choice for
$\epsilon_{eff}(y)$ has been made, it is easy to verify that a natural choice is simply
$\beta_a^{eff}(y)=\beta_ae^{[\epsilon-\epsilon_{eff}(y)]y_f}$. It follows that the total
contribution from a ``flavored" Pomeron to a Pomeron amplitude is 
\be 
 T_{a,b}(y,t)= R(y)\>
T_{a,b}^{cl}(y,t),
\ee
where $ T_{a,b}^{cl}(y,t)\equiv \beta_a\beta_be^{(1+\epsilon+\alpha'_{\cal P}t)\> y}$ is the
amplitude according to a ``high energy" description with a fixed Pomeron intercept, and $R(y)\equiv
e^{-[\epsilon-\epsilon_{eff}(y)](y-y_f)}$ is a ``flavoring" factor. In terms of  $s=e^y$, 
\be
R(s)\equiv ({s_f\over s})^{[\epsilon-\epsilon_{eff}(\log s)]}
\label{eq:FlavoringFactor}
\ee
where $s_f\equiv  e^{y_f}$. (See Figure~\ref{fig:total_xs}.) 

This flavoring factor
should  be consistently applied as part of each  ``Pomeron propagator". With  the  normalization 
 $R(\infty)=1$, we can therefore leave the residues alone, once they have been determined by a
``high energy" analysis.  For instance, for the single-particle gap  cross section before  taking
into account final-state screening, since there are three Pomeron propagators, one has for the
renormalization factor 
$Z=
R^2(y)R(y_m)
$.
It is instructive to plot in Figure~\ref{fig:flavoring}  this combination  as a function of either $\xi$ or
$M^2$ for various fixed values of $Y$.

\section{ Final Recipe}

Having explained the notion of flavoring and its effects both on Pomeron intercept and on its
residues, we need to go back to re-examine its effect for  final-state screening.
Since a Pomeron exchange enters   as a Born term, i.e., the eikonal for either  the elastic or
the inelastic diffractive production,  flavoring can easily be incorparated if we multiply both
$\chi(y,b)$ and
$\chi_d(y,b)$ by  a flavoring factor
$R(y)$. That is, if we adopt a generalized eikonal model for final-state screening, the desired
screening factor becomes 
\be
S_f(\xi,t)=S_f^{(0)}(y, t; R(y)X(y),\mu_d)
\label{eq:ScreeningFactor5}
\ee
 where $S_f^{(0)}$ is given by Eq. (\ref{eq:ScreeningFactor}). Note that  we have explicitly
exhibited here the dependence on the maximal value of the flavored elastic eikonal, $RX$,  and on the
effectiveness parameter 
$\mu_d$. 

Let's now put all the necessary ingredients together and spell out the details for our proposed
resolution to Dino's paradox.  Our {\bf final recipe} for the Pomeron contribution to
single diffraction dissociation cross section is
\be
{d\sigma \over dtd\xi}=  D_{a,{\cal P}}
(\xi, t)
\Sigma_{{\cal P}b}(M^2),
\ee
 where the unitarized flux, $D_{a,{\cal P}}$, and the Pomeron-particle cross section, $\Sigma_{{\cal
P}b}$, are given in terms of their respective classical expressions by 
$ D_{a,{\cal P}}(\xi,t)\equiv Z_d(\xi,t)F^{cl}_{a,\cal P}(\xi,t)$ and 
$\Sigma_{{\cal P}b}(M^2)\equiv Z_m(M^2)\Sigma^{cl}_{{\cal P}b}(M^2).$ It follow that the total 
suppression factor is   
\be
 Z(\xi,t;M^2)= Z_d(\xi,t)Z_m(M^2)=[S_f(\xi,t)R^2(\xi^{-1})]R(M^2),
\label{eq:TotalSuppression}
\ee
with the screening  factor given by Eq. (\ref{eq:ScreeningFactor5}) and the flavoring factor 
given by Eq. (\ref{eq:FlavoringFactor}).
Finally, we point out that the integral constraint for the unitarized flux, Eq.
(\ref{eq:SR1}), when written in terms of these suppression factors, becomes
\be
 \int_{-\infty}^0dt\int_{0}^{1}{d\xi} S_f(\xi,t)R^2(\xi^{-1})F_{a,{\cal P}}^{cl} (\xi,
t)\xi^{\epsilon} = \beta_a^{diff} <
\beta_a(0),
\label{eq:UnitarizedTPSumRule2}
\ee
where $F_{a,{\cal P}}^{cl} (\xi, t)\equiv (1/16
\pi)\beta_a(t)^2 g(t) (\xi^{-1})^{2\alpha_{\cal P}(t) -1}$.

\subsection{Phenomenological Parametrizations}

Both the  screening function and the flavoring function depend on the effective Pomeron intercept,
and  we shall adopt the following  simple parametrization.  The transition from
$\alpha^{old}(0)=1+\epsilon_o$ to
$\alpha^{new}(0)=1+\epsilon$ will occur over a  rapidity range,
$(y_f^{(1)}, y_f^{(2)})$. Let $ y_f\equiv \half(y_f^{(1)}+ y_f^{(2)})$ and $\lambda_f
^{-1}\equiv 
\half(y_f^{(2)}- y_f^{(1)})$. Similarly, we also define $\bar \epsilon\equiv
\half (\epsilon+\epsilon_o)$ and  $\Delta\equiv \half(\epsilon-\epsilon_o)$. A convenient
parametrization for $\epsilon_{eff}$ we shall adopt is 
\be
\epsilon_{eff}(y) =[\bar \epsilon +{\Delta} \tanh {{\lambda_f}
(y-\bar y_f)}].
\ee
The combination 
$[{\epsilon-\epsilon_{eff}(y)}]$ can be written as $(2{\bar\epsilon)\> [1+({s/
s_f})^{2\lambda_f}]^{-1}}$ where $ s_f=e^{ y_f}$. Combining this with Eq.
(\ref{eq:FlavoringFactor}), we arrive at a simple parametrization for our flavoring function
\be
R(s)\equiv \bigl({s_f\over s}\bigr)^{(2\bar\epsilon)\> [1+({s\over \bar s_f})^{2\lambda_f}]^{-1}}.
\label{eq:FlavoringFactor2}
\ee
With
$\alpha_{\cal P}^{old}
\simeq 1$, we have $\epsilon_o\simeq 0$, $\bar \epsilon\simeq \Delta \simeq \epsilon/2$, 
and we expect that  $\lambda_f
\simeq 1\sim 2
$ and
$ y_f \simeq 8\sim 10$ are reasonable range for these
parameters.~\cite{GlobalFlavoring}

To complete the specification, we need to provide a more phenomenological description for the
final-state screening factor. First, we shall  approximate the screening factor by an exponential in
$t$:
\be
S_f(y,t)\simeq S_f(y,0)e^{ \Delta B_d(y) t}.
\label {eq:FinalScreeningFactor}
\ee 
where $S_f(y,0)=\{r\>x^{-r}\Gamma(x,r)\}^2,$ with ${x=\mu_dR(y)X(y)}$ and $r=r(y)$. The width,
$\Delta B_d(y)$, can be obtained by a corresponding substitution.  Note that  $S_f(y,0)$ depends on
$B(y)$,
$B_d(y)$,
$X(y)$, and 
$\mu_d$.  Phenomenological studies allow us to
approximate $B(y)\simeq b_0 + .25 y$ and $B_d(y)\simeq b_d + .25 y$, $b_d\simeq b_0/2\sim 2.3\>
GeV^{-2}$. 

The only
quantity left to be specified is the effectiveness parameter $\mu_d$. Since the physics of
final-state screening is that driven by a Pomeron with intercept greater than unity, the relevant
rapidity scale is again
$y_f$. Let us  fix
$\mu_d$ first by requiring that  screening is small for 
$y<y_f$, {\it i.e.}, 
$S_f(y,0)\sim 1$ as one moves down in rapidity from $y_f$ to  $ y_r$. Similarly, we
expect screening to approach its full strength as one moves past the flavoring threshold $ y_f$.
We thus find it economical to parametrize 
\be
\mu_d(y)\simeq (\mu_0/ 2) \{1+ \tanh [{\lambda_d (y-\bar y_f)}]\},
\ee
and we expect $\lambda_{d}\sim \lambda_f$.   This completes the specification of
our unitarization procedure.

\subsection{Qualitative Discussion}

Before attempting to {``\bf use"} our recipe, it is important to gain a qualitative  understanding
of what one can expect.  Note that the sole effect of the flavoring factor
$R$ is to ``restore" a Pomeron contribution to that of a ``bare Pomeron" with an intercept
$\alpha^{old}_{\cal P}\simeq 1$. It follows that  $R(y) \sim 1$ for
$y>>y_f$ and $R(y)>1$ for $y<<y_f$. Similarly, since we expect final-state screening to increase as
$\epsilon_{eff}(y)\sim \epsilon$, we therefore expect $S_f(y,t)< 1$ for $y>>y_f$ and
$S_f(y,t)\rightarrow 1$ for $y<<y_f$.

Let us examine the phase
space region for a singly diffraction dissociation process from ISR to  LHC. (See Figure~\ref{fig:phase_space}.) 
  It is customary to adopt $y\geq 3$ and  ${y_m}\geq 0.4$ for a triple-Pmeron
analysis, (corresponding to $\xi_{max}\sim 0.05$ and $\xi_{min}=1.5/s$.) However, a triple-Pomeron
analysis cannot be reliably used for
$0<y_m\leq y_r$. To be conservative, we adopt the same lower cutoff
$y_{min}\simeq 3$ for both $y$ and $y_m$. With
$y_f\simeq 8\sim 10$,  the phase space is divided into the following four  main regions of interest:
\begin{itemize}
\item 
(A) $y_{min}<y_m<y_f$, $y_{min}<y<y_f$: 

$\alpha_{eff}(y)\simeq 1$, $\alpha_{eff}(y_m)\simeq 1$, and  $d\sigma^{sd}/dy \sim constant$.
\item
(B) $y_m<y_f$,  $y_{min}<y<y_f$: 

$\alpha_{eff}(y)\simeq 1$, $\alpha_{eff}(y_m)\simeq
1+\epsilon$, and  $d\sigma^{sd}/dy \sim e^{\epsilon y_m}=(M^2)^{\epsilon}$. 
\item
(C) $y_{min}<y_m<y_f$, $y>y_{f}$: 

$\alpha_{eff}(y)\simeq 1+\epsilon$,
$\alpha_{eff}(y_m)\simeq 1$, and  $d\sigma^{sd}/dy \sim S_f(y,t) e^{2\epsilon
y}=S_f(\xi,t)\xi^{-2\epsilon}$.  
\item 
(D) $y_m>y_f$,  $y>y_f$: 

$\alpha_{eff}(y)\simeq 1+\epsilon $, $\alpha_{eff}(y_m)\simeq
1+\epsilon$, and $d\sigma^{sd}/dy \sim S_f(y,t) e^{2\epsilon
y}e^{\epsilon y_m}=S_f(\xi,t)\xi^{-\epsilon}s^{\epsilon}.$

\end{itemize}
In region-(A),  diffractive cross sections behave according to the ``old-fashion"
Triple-Pomeron formula, with $\alpha_{\cal P}(0)=1$. 
 Region-(B) is perhaps  the best place  where one has a clean  chance of a direct detection of
the $\epsilon>0$ effect  coming from the
$M^2$-dependence at fixed
$y<y_f$. 
One would normally also expect to be able to detect the $\epsilon>0$ effect  directly in
Region-(C). However, in the large-$y$ limit where the strength of final-state screening   becomes
important,the suppression from $S_f(y,t)$ will overcome  the $e^{2\epsilon y}$ behavior.
Region-(D) exits only if $Y>2y_f$.  With
$y_f\simeq 8\sim 10$, this region becomes  interesting only as one moves from  Tevatron to LHC
energies.

Consider the two available energies at Tevatron. Moving along a fixed $Y$ line, depending on the
strength of screening, the cross section could be  suppressed in the region-(C), and  most of
the contribution to
$\sigma^{sd}$ will come from region-(A) and region-(B).  As one moves
from $\sqrt s=546 \> GeV$ to $\sqrt s= 1800\> GeV$, the rapidity span in  region-B decreases.
In contrast, below Tevatron energies, one
is most likely to be in region-(A). Even more strikingly, at these ``low energies",  a
large portion of the kinematic region lies in regions where either $y$ or $y_m$ is near  $y_r$ or
below. As a consequence, we would not be surprised if our triple-Pomeron formula turns out to
lead to a $\sigma^{sd}$  which is lower than what has been measured; other triple-Regge
contributions should be included.

\subsection{ A Caricature of High Energy Diffractive Dissociation}

The most important new parameter we have introduced for understanding high energy diffractive
production is the flavoring scale, $s_f=e^{y_f}$. We have motivated by way of a simple model to
show that a reasonable range for this scale is $y_f\simeq 8\sim 10$. Quite independent of
our estimate, it is possible to treat our proposed resolution phenomenologically and determine this
flavoring scale from experimental data.

It should be clear that
one is  not attempting to carry out a full-blown phenomenological analysis here. To do that, one must
properly incorporate other triple-Regge contributions, {\it e.g.}, the
${\cal PPR}$-term for the low-$y_m$ region, the $\pi\pi{\cal P}$-term and/or the ${\cal RRP}$-term 
for the low-$y$ region, etc., particularly for $\sqrt s \leq \sqrt {s_f}\sim 100\> GeV$.  What we 
hope to achieve is to provide a ``caricuture" of the interesting physics involved in diffractive
production at collider energies through our introduction of the screening and the flavoring
factors.~\cite{GlobalFlavoring}   

 Let us begin by first examining  what we should  expect. Concentrate on  the triple-Pomeron vertex
$g(0)$ measured at high energies. Let us for the moment assume that it has also been meassured
reliably at low energies, and let us denote it as
$g^{low}(0)$. Our flavoring analysis  indicates that these two couplings are related by
\be
g(0)\simeq e^{-({3\epsilon y_f\over 2})}g^{low}(0).
\label{eq:FlavoringTPCoupling}
\ee
With $\epsilon\simeq 0.08\sim  0.1$ and $y_f\simeq 8\sim 10$,
using the value $g^{low}(0)=0.364\pm 0.025\>\> mb^{\half}$,~\cite{Dino5} we expect a value of
$0.12\sim 0.18\> mb^{\half}$. Denoting the overall multiplicative constant for our renormalized
triple-Pomeron formula by
$K$,
\be 
K\equiv {\beta^2_a(0)g_{\cal PPP}(0)\beta_b(0)\over 16\pi}.
\ee
With $\beta^2_p\simeq 16\> mb$, we therefore expect $K$ to lie between the range $.15\sim .25\>\>
mb^2$.

We begin  testing our renormalized triple-Pomeron formula by first turning off  the
final-state screening, {\it i.e.,} setting $S_f=1$.  We determine the overall multiplicative
constant $K$ by normalizing the integrated $\sigma^{sd}$ to the measured CDF $\sqrt s=1800\> GeV$
value.\cite{CDF} With $\epsilon=0.1$,
$\lambda_f=1$, this is done for a series of values for $y_f=7,\>8,\>9,\>10$. We obtain 
respective values for $K=.24,\> 0.21,\> 0.18,\> 0.15,$ consistent with our flavoring expectation.
As a further check on
the sensibility of these values for the flavoring scales, we find for the ratio $\rho\equiv
\sigma^{sd}(546)/\sigma^{sd}(1800)$ the values $0.63, \> 0.65,\> 0.68,\> 0.72$ respectively. This
should be compared with the CDF result of 0.834.

Next we consider screening. Note that screening would increase our values for $K$, which would lead
to large values for $g$. Since we have already obtained values for triple-Pomeron coupling which are
of the correct order of magnitude, the only conclusion we can draw is that, at Tevatron, screening 
cannot be too large. With our parametrization, we  find that screening is rather small at Tevatron
energies, with $\mu_0\simeq 0.0\sim 0.2$ This comes as somewhat a surprise! Clearly, screening will
become important eventually at higher energies. After flavoring, the amount of screening required at
Tevatron is apparently greatly reduced.

Having shown that our renormalized triple-Pomeron formula does lead to sensible predictions for
$\sigma^{sd}$ at Tevatron, we can improve the fit by enhancing the $PPR$-term as well as
$RRP$-terms which can become important. Instead of introducing a more involved phenomenological
analysis,  we simulate the desired low energy effect by having
$\epsilon_o\simeq -0.06\sim -0.08$. A remarkably good fit results with 
$\epsilon=0.08\sim 0.09$, $y_f=9$ and $\mu_0\simeq 0\sim 0.2$.~\cite{GlobalFlavoring} This is shown
in Figure~\ref{fig:diff_xs}. The ratio
$\rho$ ranges from $0.78\sim 0.90$, which is quite reasonable. The prediction for $\sigma^{sd}$ at
LHC is $12.6\sim 14.8\> md$.

Our fit leads
to a triple-Pomeron coupling in the range of 
\be
g_{\cal PPP}(0)\simeq .12\sim .18 \>\>{mb}^{\half}, 
\ee
exactly as expected. Interestingly, the triple-Pomeron coupling quoted in Ref. \cite{Dino1}
($g(0)=0.69
\>mb^{1\over 2}$) is actually a factor of 2 larger than the corresponding low energy
value.~\cite{Dino5} Note that this difference  of a factor of 5 correlates almost precisely with the
flux renormalization factor 
$N(s)\simeq 5$ at Tevatron energies.

We believe, with care, the physics
of flavoring and final-state screening can be tested independent of the specific
parametrizations we have proposed here. In particular, because our unitarized Pomeron flux approach
retains factorization along the ``missing mass" link, unambiguous predictions can be made for other
processes involving rapidity gaps.

\section {Predictions for  Other Gap Cross Sections}

 For both double Pomeron exchange (DPE)
and doubly diffractive (DD) processes, one is dealing with three rapidity variables which can
become large. We will treat these two cases first  before turning to more general situations.

\noindent{\bf Prediction for DPE  Cross Sections:}

For double Pomeron exchange (DPE), we are dealing with events with two large rapidity gaps. The
final state configuration can be specified by five variables, $t_1$, $t_2$, $\xi_1$, $\xi_2$, and
$M^2$. For $t_1$ and $t_2$ small, one again has a constraint, $\xi_1^{-1}M^2\xi_2^{-1}\simeq s$.
Alternatively, we can work with rapidity variables, $y_1\equiv \log (\xi_1^{-1})$, $y_2\equiv \log
(\xi_2^{-1})$, and $y_m\equiv \log M^2$, with $y_1+y_m+y_2\simeq Y=\log s$. The appropriate  DPE
differential cross section can be written down, with no new free parameter. 
Let us introduce a renormalization  factor 
\be 
{d\sigma\over dy_1dt_1dy_2dt_2}= Z_{DPE}(y_1,t_1,y_m,y_2,t_2){d\sigma^{classical}\over
dy_1dt_1dy_2dt_2}.
\ee
One immediately finds that, using Pomeron factorization for the missing mass variable,  
\be
Z_{DPE}=[S_f(y_1,t_1)R(y_1)^2]R(y_m)[S_f(y_2,t_2)R(y_2)^2]
=Z_d(\xi_1,t_1)Z_m(M^2)Z_d(\xi_2,t_2).
\ee

Alternatively, we can express this cross section in terms of singly diffractive dissociation cross
sections as
\be
{d\sigma_{ab}\over dy_1dt_1dy_2dt_2}=\{{d\sigma_{ab}\over
dy_1dt_1}\}\{R(y_m)\sigma^{cl}_{{ba}}(y_m)\}^{-1} \{{d\sigma_{ab}\over dy_2dt_2} \}
\ee
where $\sigma^{cl}_{{ba}}(y_m)=\beta_{b}\beta_{a}e^{\epsilon\>y_m}$. This {\bf clean} prediction
involves no new parameter, with the understanding that, when $y_m$ is low, secondary
terms must be added.

\noindent{\bf Prediction for  DD Cross Sections:}

For double diffraction dissociation (DD), there are two large missing mass variables, $M_1^2$,
$M_2^2$, separated by one large rapidity gap, $y$,  and its associate momentum transfer variable
$t$. Again, for $t$ small, we have the constraint $y_{m_1}+y_{m_2}+y\simeq Y$. 

The classical differential DD cross section is
$\Sigma^{cl}_{a,{\cal P}} (y_{m_1}) {\bar F}^{cl}_{{\cal P}} (y,t)\> 
\Sigma^{cl}_{b,{\cal P}} (y_{m_2}),   
$
where the  classical ``gap distribution" function is $\bar F^{cl}_{\cal P}(t, y)= (1/16 \pi)
e^{2(\epsilon+\alpha'_{\cal P}t) y}g^2_{\cal PPP}(t).$
After taking care of both flavoring and final-state screening, on obtains for the renormalization
factor
\be
Z_{DD}(y_{m_1},y,t, y_{m_2})=R(M_1^2)[\bar S_f(y,t)R(y)^2] R(M^2)\equiv Z_m(M_1^2)\bar
Z_d(\xi,t)Z_m(M^2_2).
\ee
A new  screening factor, $\bar S_f(y,t)$,  has to be introduced because of the difference in the
$t$-distribution associated with two factors of triple-Pomeron coupling.   It can be obtained from
$S_f(y,t)$ by replacing $B_d(y)$  by $\bar B(y)=\bar b+\alpha_{\cal P}' t$,
where, by factorization, 
$\bar b=2b_d-b_0$, ($\bar b$ is the t-slope associated with the triple-Pomeron coupling). 

Alternatively, this cross section can again be expressed as a product of two single diffractive
cross sections
\be
{d\sigma_{ab}\over
dy_{m_1}dt dy_{m_2}}= \{{d\sigma_{ab}\over
dy_{m_1}dt}\}\{\bar Z_d(y,t)\sigma^{el}_{ab} (y,t)\}^{-1}  \{{d\sigma_{ab}\over
dy_{m_2}dt}\},
\ee
where $\sigma^{el}_{ab} (y,t)\equiv (1/16 \pi)|\beta_a(t)\beta_b(t)
e^{(\epsilon+\alpha'_{\cal P}t) y}|^2.$
Other than the modification from $S_f$ to $\bar S_f$, this prediction is again given uniquely in
terms of the single diffraction dissociation cross sections.

\noindent{\bf Other Gap Cross Sections}

We are now in the position to write down the general Pomeron contribution to the differential cross
section with an arbitrary number of large rapidity gaps. For instance, generalizing the DPE process
to  an
$n$-Pomeron exchange process, there will now be  $n$ large rapidity gaps, with $n-1$ short-range
ordered missing mass distributions alternating between two gaps. The corresponding renormalization 
factor is
\be
 Z^{(n)}_{PE}=Z_d(\xi_1,t_1)Z_m(M_1^2)\bar Z_d(\xi_2,t_2) Z_m(M^2_2)  \cdots\cdots
Z_m(M_{(n-1)}^2) Z_d(\xi_n,t_n).
\ee
Other generalizations are all straight forward.
However, since these will unlikely be meanful phenomenologically in the near future, we shall not
discuss them here.  It is
nevertheless interesting to point out that, if any cross section does become meaningful
experimentally, flavoring would dictate that it is most likely the classical triple-Regge formulas
with
$\alpha_{\cal P}(0)\simeq 1$ that would be at work first.

\section{ Final Remarks:}

Let us briefly recapitulate what we have accomplished.  Given Pomeron as a pole, the total Pomeron
contribution to  a singly diffractive dissociation cross section can in principle be expressed as
\be
{d\sigma \over dtd\xi}=[S_i(s,t)][D_{a,{\cal P}}(\xi,t)] [\Sigma_{{\cal P} b} (M^2)],
\label{eq:UnitarizedTP}
\ee
\be
D_{a,{\cal P}} (\xi, t)= S_f(\xi,t)F_{{\cal P}/a} (\xi, t)
\label{eq:UnitarizedPFlux}
\ee
\begin{itemize}
\item
 The first term, $S_i$, represents initial-state screening correction.  We have  demonstrated that,
with a Pomeron intercept greater than unity and with  a pole approximation for total and elastic
cross sections remaining valid, initial-state absorption cannot be large.  We therefore can justify 
setting 
$S_i\simeq 1$ at Tevatron energies.  

\item The first crucial step in our alternative resolution to the
Dino's paradox lies in properly treating the final-state screening, $S_f(\xi,t)$.
We  have explained in an expanding disk setting why a final-state screening can set in relatively 
early when compared with that for elastic and total cross sections.

\item
 We have stressed that the dynamics of a soft Pomeron in a
non-perturbative QCD scheme requires taking into account the effect of ``flavoring", the notion that
the effective degrees of freedom for Pomeron is suppressed at low energies. As a consequence, we
find that 
$ F_{{\cal P}/a} (\xi, t)=R^2(\xi^{-1}) F^{cl}_{{\cal P}/a} (\xi, t)$ and  $ \Sigma_{{\cal
P}b}(M^2)=R(M^2)\Sigma_{{\cal P}b}^{cl}(M^2)$ where $R$ is a ``flavoring" factor. 
\end{itemize}

It is perhaps worth  contrasting what we have achieved with the flux renormalization
scheme of Goulianos.~\cite{Dino1} By construction, the normalization factor $N(s)$ is of the form
which one would have obtained from   an initial-state screening consideration. Although this breaks
factorization, one might hope perhaps the scheme could be phenomenologically
meaningful at Tevatron energies. Note that, for
$\sqrt s>22$ $ GeV$, the renormalization factor $N(s)$ has an approximately factorizable form: 
$N(s)\sim .25\>s^{2\epsilon}=.25\> (\xi^{-1})^{2\epsilon}(M^2)^{2\epsilon}$.  it follows that the
diffractive differential cross section remains in   a factorized form: 
\be
\xi
{d\sigma^{sd}\over dtd\xi}\sim .25 [\xi^{2\epsilon+1}F_{{\cal P}/p}^{(0)} (\xi, t)]
[(M^2)^{-2\epsilon}\sigma_{{\cal P}p}^{(0)} (M^2,t)].
\label {eq:DinoDistribution}
\ee
 It can
be shown that Eq. (\ref{eq:DinoDistribution}) leads to a diffractive cross section $\sigma^{sd}$
which, up to $\log s$,  is asymptotically constant. That is, the diffractive dissociation
contribution no longer  corresponds to the part of total cross sections represented by the Pomeron
exchange. This   is not in accord with the basic hypothesis of Pomeron dominance for total and
elastic cross sections at Tevatron energies.~\cite{Dino8}

Our final resolution shares certain common features with that proposed by Schlein.~\cite{Schlein1}
 At a fixed $\xi$, 
$Z_m(M^2)\simeq 1$ as $s\rightarrow \infty$ so that it is  possible to identify our renormalization
factor
$Z_d(\xi,t)=S_f(\xi,t)R^2(\xi^{-1})$ with the flux damping factor $Z_{S}(\xi)$ of Schlein. In Ref.
\cite{Schlein1}, it was emphasized that the behavior of $Z_S(\xi)$ can be separated into three
regions. (i) $(\xi_1,  \xi_{max})$ where $Z_S\simeq 1$, (ii) $(\xi_2, \xi_1)$ where $Z_S$ drops
from 1 to 0.4 smoothly,  and (iii) $(0,\xi_2)$ where  $Z_S(\xi) \rightarrow 0$ rapidly as
$\xi\rightarrow 0$.  The boundaries of these regions are $\xi_1\sim 0.015$ and $\xi_2\sim
10^{-4}$. The first bounday $\xi_1$ can be identified with our energy scale, $s_r$, 
$\xi_1^{-1} \sim y_r$.  If we identify the  boundary between region-(ii) and
region-(iii) with our flavoring scale $y_f$ by 
$s_f^{-1}=e^{-y_f}=\xi_2$, one has
$y_f\simeq 9$, which is consistent with our estimate. Since $S_f(\xi,t)\simeq 1$ for
$\xi>s_f^{-1}$ and $R^2(\xi^{-1})$ drops from $R^2(1)\simeq s_f^{2\epsilon}$ to 1 at $s_f$, their
$Z_S(\xi)$ behaves qualitatively like our renormalization factor. If one indeed makes this
connection, what had originally been a mystery for the origin of the scale, $\xi_2$,  can now be
related to the non-perturbative dynamics of Pomeron flavoring.~\cite{Schlein8}

It should be stressed  that our discussion depends crucially on the notion of  soft Pomeron
being a factorizable Regge pole. This notion  has always  been controversial.
Introduced more than thirty years ago, Pomeron was identified as the leading Regge trajectory with
quantum numbers of the vacuum with
$\alpha(0)\simeq 1$ in order to account for the near constancy of the low energy hadronic total cross
sections. However, as a Regge trajectory, it was unlike others which can be identified by the
particles they interpolate. With the advent of QCD, the situation has improved, at least
conceptually. Through
large-$N_c$ analyses and through other non-perturbative studies, it is natural to expect
Regge trajectories in QCD as manifestations of ``string-like" excitations for bound states and
resonances of quarks and gluons due to their long-range confining forces. Whereas ordinary meson 
trajectories can be thought of as ``open strings" interpolating $q\bar q$ bound states, 
Pomeron corresponds to  a ``closed string" configuration associated with glueballs. However, the
difficulty of identification, presummably due to strong mixing with  multi-quark states,
has not helped the situation in practice. In a simplified one-dimensional multiperipheral
realization of large-N QCD, the non-Abelian gauge nature nevertheless managed to re-emerge
through its topological structure.~\cite{LeeVeneziano}\cite{DPM}

 The
observation of ``pole dominance" at collider energies has hastened the need to examine more
closely various  assumptions made for Regge hypothesis from a more fundamental viewpoint. It is
our hope that by  examining Dino's paradox carefully and by  finding an alternative resolution to the
problem without deviating drastically from accepted guiding principles for hadron dynamics, Pomeron
can continued to  be understood as a Regge pole in a non-perturbative QCD setting. The resolution
for this paradox could therefore lead to a re-examination of other interesting questions from a
firmer theoretical basis.  For instance,  to be able to relate quantities such as the  Pomeron
intercept  to non-perturbative physics of color confinement represents a theoretical challenge of
great importance.

{\bf Acknowledgments:}
I would like to thank  K. Goulianos for first getting me interested in this problem during the
Aspen Workshop on Non-perturbative QCD, June 1996.  Intensive   discussions with K. Goulianos, A.
Capella, and A. Kaidalov at  Rencontres de Moriond, March, 1997, have been extremely helpful. I
am also grateful to  P. Schlein for explaining to me details of their work and for his advice.
 I want to thank both K. Goulianos and P. Schlein for   helping  me to understand what
I should or should not believe in various facets of diffractive data!  Lastly, I really appreciate
the help from K. Orginos for both  numerical analysis and the preparation for
the figures. This work is supported in part  by the D.O.E.  Grant
\#DE-FG02-91ER400688, Task A.

\newpage

\begin{figure}
$$
\epsfxsize=\textwidth
\epsfysize=\textwidth
\epsfbox{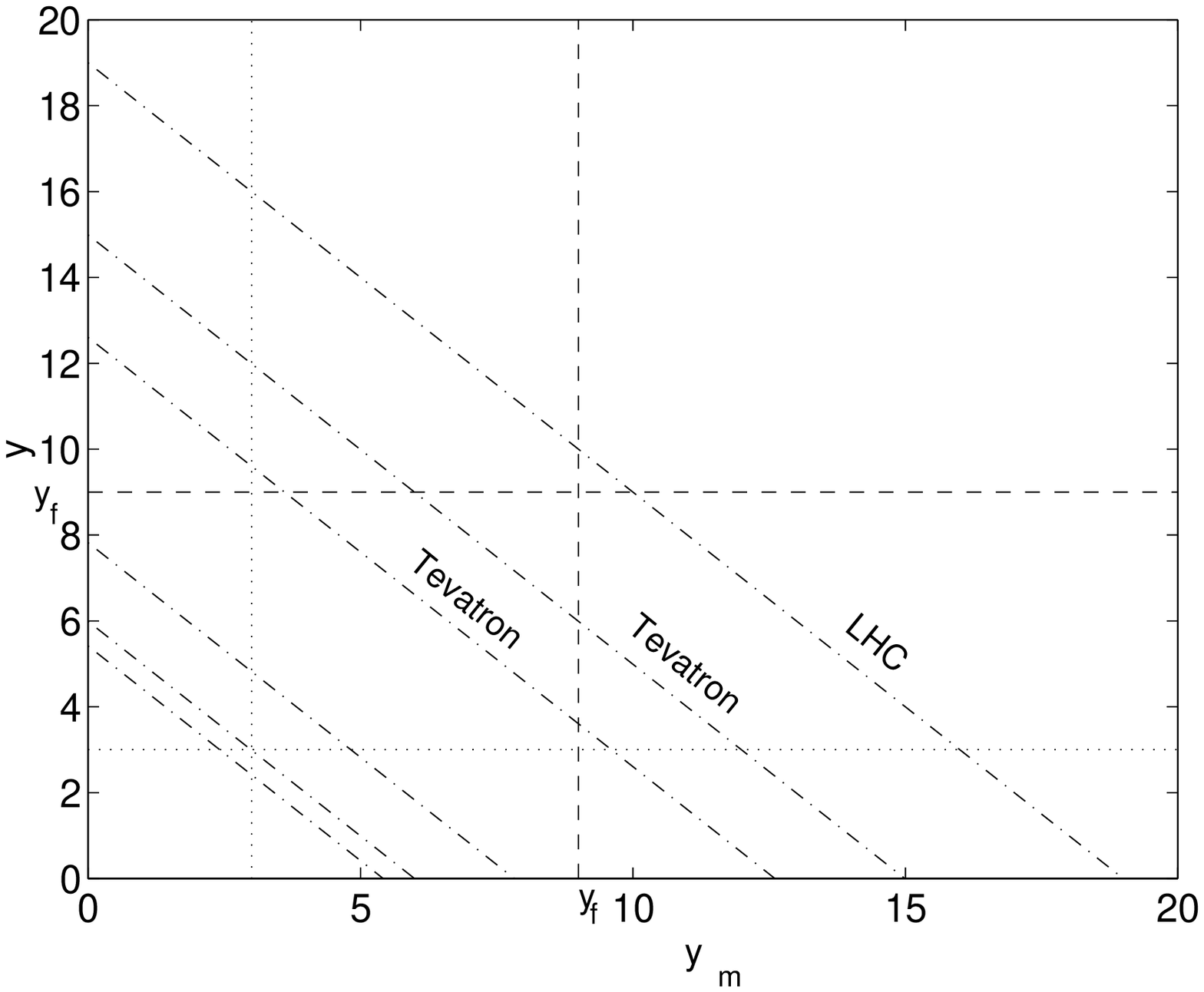} 
$$
\vspace{.1cm }
\caption{Phase space for single diffraction dissociation from ISR to LHC in terms of rapidity
variables
$y\equiv \log {\xi^{-1}}$ and $y_m\equiv \log M^2$. The dashed lines are for ``flavoring"
scale, $y_f$, chosen to be 9 for illustration. Dotted lines are  $y_{min}\simeq 3$ lines for
both $y$ and $y_m$. The Dashed-dotted lines are constant center of mass energy lines for
{$E_i$},
$i=1,\cdots\cdots, 6,$ equals to  $15, 30, 60, 546, 1800, 14,000\>GeV$ respectively.}
\label{fig:phase_space} 
\end{figure}

\begin{figure}
$$
\epsfxsize=\textwidth
\epsfbox{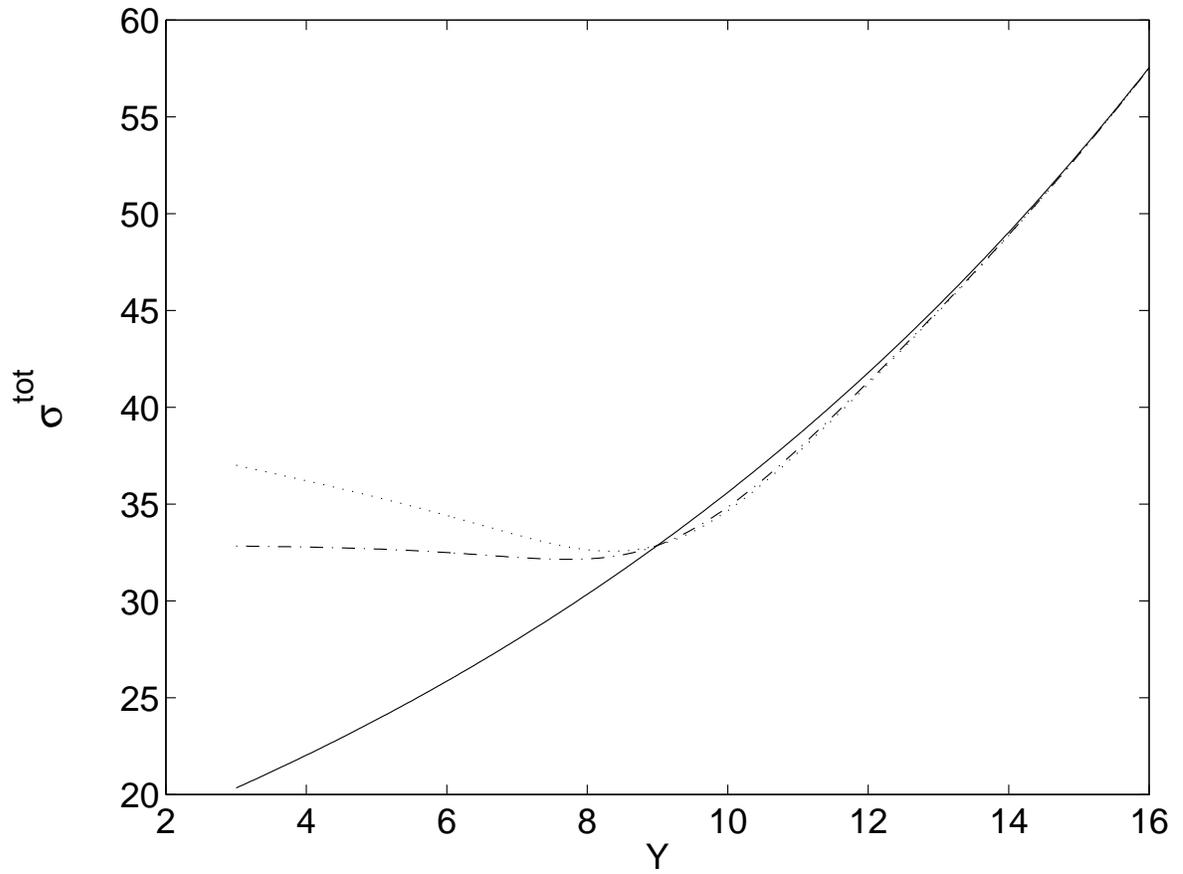} 
$$
\vspace{.1cm }
\caption{Effect of flavoring factor $R(s)$ when applied to a standard  rising cross
section: $\sigma^{cl}=\beta^2\> s^{\epsilon}$,  $\epsilon=0.1$ and  $\beta^2=16\> mb$, given by the
solid curve. The 
 dashed-dotted curve has   $\epsilon_o=0$, 
$\lambda_f=1$, and flavoring scale $y_f=9$. The dotted curve
corresponds to 
$\epsilon_o=-0.04$.} 
\label{fig:total_xs} 
\end{figure}

\begin{figure}
$$
\epsfxsize=\textwidth
\epsfbox{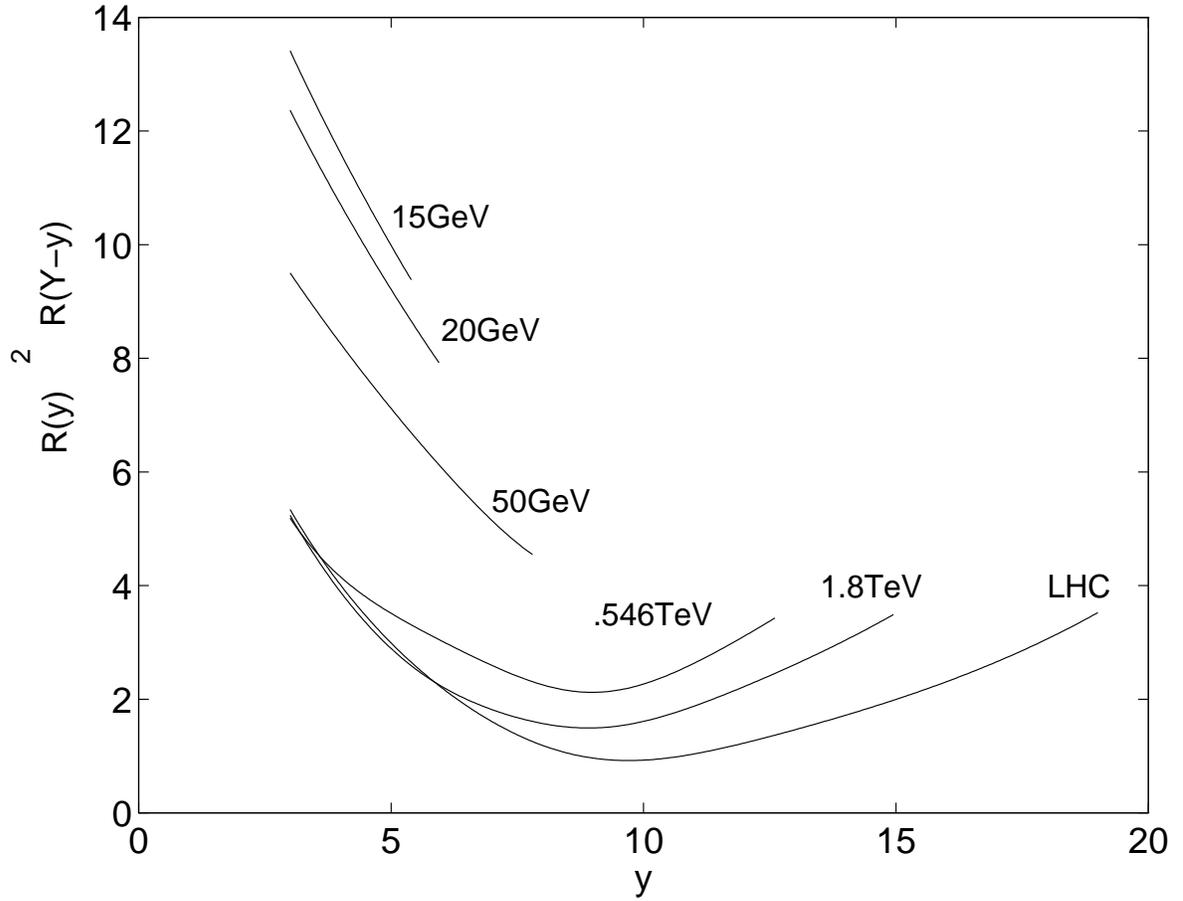} 
$$
\vspace{.1cm }
\caption{Renormalization factor due to flavoring alone, $Z_f(\xi;s)\equiv R^2(\xi^{-1})R(M^2)$,
as a function of rapidity $y=\log \xi^{-1}$ for various fixed center of mass energies.}  
\label{fig:flavoring} 
\end{figure}

\begin{figure}
$$
\epsfxsize=\textwidth
\epsfbox{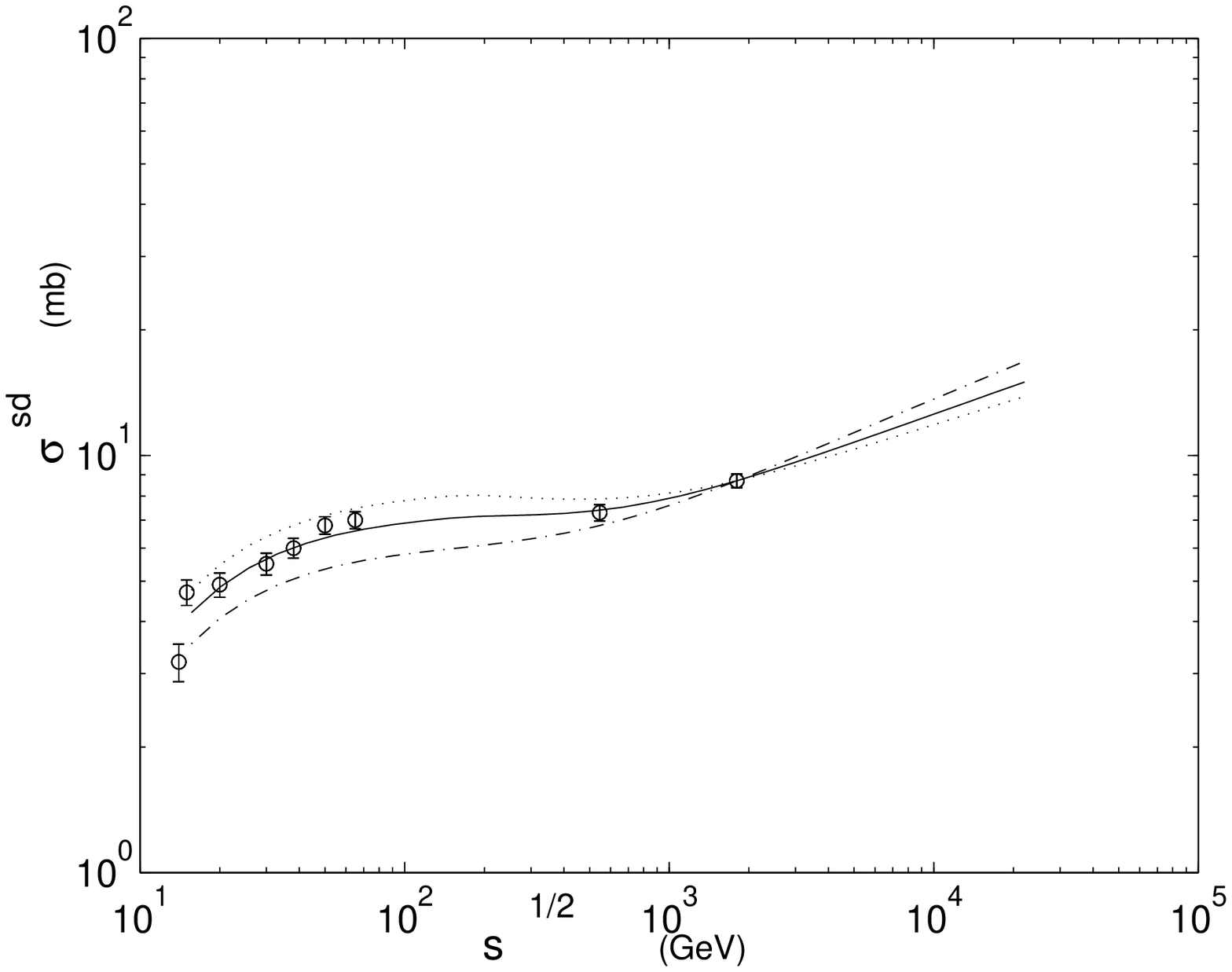} 
$$
\vspace{.1cm }
\caption{Various fits to representative single diffraction dissociation cross sections extracted
from Ref. \cite{Dino1} from ISR to Tevatron. The solid line corresponds to    $\epsilon=0.08$,
$\epsilon_o=-0.07$, 
$\lambda_f=1$,  $y_f=9$, with a small amount of  final-state screening, $\mu_0=0.1$.  The
dotted curve corresponds to $\mu_0=0.2$, and the dashed-dotted curve corresponds to no
screening, $\mu_0=0$.}
\label{fig:diff_xs} 
\end{figure}


\begin{thebibliography}{99}
\bibitem{Convention}{When compared with
the classical  formula, our decomposition corresponds to having the triple-Regge
coupling absorbed into the
 flux factor, $D_k$. For our
subsequent discussion, the triple-Pomeron coulping $g(t)$ will be absorbed  into the classical flux
factor $F_{{\cal P}/a}^{cl} (\xi, t)$ and the Pomeron-particle cross section without  $g(t)$
 will be denoted by $\Sigma_{{\cal P}b}^{cl}(M^2)$.}
\bibitem{Diffraction}{ For diffraction dissociation,
$c$ must have same quantum numbers as the incident particle $a$. For
definiteness, we   discuss leading particle gap distribution relative to the  incoming
particle
$a$ with a positive, large rapidity, and assume $c=a$.  For $p\bar p$ collision, actual diffractive
cross section,
$\sigma^{sd}$,  is arrived at by taking into account  contributions involving diffraction at both
$p$ and $\bar p$ vertices.}
\bibitem{EdgeEffect} {This mechanism has been recognized before. It was used in Refs.~
\cite{Flavoring5} and \cite{Odderon} to explain why a ``maximal odderon" cannot be allowed in any
hadronic scheme which admits an  expanding-disk interpretation at high energies.}
\bibitem{InitialState} {It is
possible also to apply an eikonal model to study initial state screen for singly
diffractive dissociation cross section. This indeed has been performed before,
however, without taking final-state screening into account.~\cite{GotsmanLevinMaor}
The fact that inelastic absorption takes
place at small impact parameter, with surviving scattering allowed only at the edge
of the expanding disk has also been noted there. Since our final-state absorption would remove all
scattering within the disk, applying an eikonal  initial-state absorption procedure becomes
unnecessary.}
\bibitem{Flavoring6} We do not include ``semi-hard" production in the current treatment for soft
Pomeron. Flavoring will indeed be the primary mechanism in our construction of a ``Heterotic
Pomeron".  As mentioned in Ref.
\cite{Tan1}, this will be treated in a separate publication.
\bibitem{GlobalFlavoring} By choosing $\epsilon^{old}< 0
$, it is possible to provide a global ``average" description mimicking ``secondary trajectory"
contributions for various low energy regions. In Ref. \cite{Flavoring1},  acceptable estimates are
$\epsilon_o\simeq -0.11\sim -0.5$.
\bibitem{Schlein8} There are also several   differences between our result and
that of Schlein. First, our renormalization factor $Z(\xi,t;s)$ is t-dependent whereas Schlein's is
not. At very small $\xi$, our suppression factor does not vanish as fast as that of
Schlein:  
$\bigl($ $Z_S(\xi)\sim \xi$, whereas ours behaves as $\xi^{2\epsilon}$ $\bigr)$. Furthermore, since
we have found that there is very little screening needed at Tevatron energies, our slow cutoff might
not set in until much higher energies so that it could  indeed be possible to observe the
$\xi^{-2\epsilon}$ behavior for $d\sigma^{sd}/dydt$ in Region-(C) at Tevatron energies.
\end{thebibliography}
\end{document}